\documentstyle[prl,aps]{revtex}

\input epsf
\begin{document}
\draft
\twocolumn[\hsize\textwidth\columnwidth\hsize\csname@twocolumnfalse\endcsname

\preprint{} 
\title{Superconducting Gap Structure of
$\kappa$-(BEDT-TTF)$_2$Cu(NCS)$_2$ Probed by Thermal Conductivity
Tensor}

\author{K.~Izawa$^{1}$, H.~Yamaguchi$^{1}$, T.~Sasaki$^{2}$, and
Yuji~Matsuda$^{1}$}
\address{$^1$Institute for Solid State Physics, University of Tokyo, 
Kashiwa, Chiba 277-8581, Japan}
\address{$^2$Institute for materials research, Tohoku University, 
Sendai 980-8577, Japan}
\maketitle

\begin{abstract}

The thermal conductivity of organic superconductor
$\kappa$-(BEDT-TTF)$_2$Cu(NCS)$_2$ ($T_c$=10.4~K) has been studied in
a magnetic field rotating within the 2D superconducting planes with
high alignment precision.  At low temperatures ($T \alt 0.5$~K), a
clear fourfold symmetry in the angular variation, which is
characteristic of a $d$-wave superconducting gap with nodes along the
directions rotated 45$^{\circ}$ relative to the {\boldmath $b$} and
{\boldmath $c$} axes of the crystal, was resolved.  The determined
nodal structure is inconsistent with recent theoretical predictions of
superconductivity induced by the antiferromagnetic spin fluctuation.
	
\end{abstract}

]

\narrowtext
Since the discovery of superconductivity in organic materials about 2
decades ago, the question of the pairing symmetry among this class of
materials is one of the most intriguing problems.  In particular, the
nature of the superconductivity in quasi-2D $\kappa$-(BEDT-TTF)$_2$X
salts ($\kappa$-(ET)$_2$X), where the ion X can, for example, be
Cu(SCN)$_2$, Cu[N(CN)$_2$]Br or I$_3$, has attracted considerable
attention.  In these layered organics, Shubnikov-de Haas oscillation
experiments have established the existence of a well-defined Fermi
surface (FS), demonstrating the Fermi liquid character of the low 
energy
excitation.  The large enhancement of the effective mass revealed by
the specific heat as well as magnetic susceptibility measurements
suggests the strong electron correlation effect in the normal state. 
Moreover, it was suggested that superconductivity occurs in proximity
to the antiferromagnetic (AF) ordered state in the phase diagram
\cite{kanoda}.  Since some of these unusual properties suggest
analogies with high-$T_c$ cuprates \cite{mckenzie1}, it was pointed
out by many authors that the AF spin-fluctuation should play an
important role for the occurrence of superconductivity
\cite{AFSF1,AFSF2}.

Unconventional superconductivity is characterized by a superconducting
gap with nodes along certain crystal directions.  Since the
superconducting gap structure is intimately related to the pairing
interaction, its identification is crucial for understanding the
pairing mechanism.  Although the structure of the superconducting
order parameter of $\kappa$-(ET)$_2$X salts has been examined by
several techniques, it is still controversial as we now summarize
\cite{kanoda}.  Results strongly in favor of unconventional pairing
symmetry came from NMR experiments of
$\kappa$-(ET)$_2$Cu[N(CN)$_2$]Br, in which the absence of the
Hebel-Slichter peak and cubic $T$-dependence of the spin lattice
relaxation rate $1/T_1$ were interpreted as an indication of $d$-wave
pairing with line nodes \cite{kanoda,jerome}.  The existence of the
$T$-linear term in the thermal conductivity at low temperatures of
$\kappa$-(ET)$_2$Cu(NCS)$_2$ also supported the presence of line nodes
\cite{belin}.  However, some of the specific heat and penetration
depth studies on these materials led to conflicting results.  For
example, recent specific heat measurements reported a fully gapped
superconductivity \cite{elsinger}.  Since these measurements rely on
the $T$-dependence of the physical quantities, it is more desirable to
measure the in-plane anisotropy of the gap directly in order to probe
the gap structure.  Very recently, such an attempt was made by STM
\cite{nomura} and mm-wave transmission \cite{schrama} experiments. 
Although both measurements reported the strong modulation of the gap
structure, they led to completely different conclusions on the node
directions; the former predicts nodes along the directions rotated
45$^{\circ}$ relative to the {\boldmath $b$} and {\boldmath $c$}-axes
while the latter predicts nodes along the {\boldmath $b$} and
{\boldmath $c$} directions.  In interpreting these experiments, one
needs to bear in mind that the STM spectrum parallel to the 2D plane
can be strongly affected by the atomic state at the edge.  Moreover,
an alternative interpretation was proposed for the mm-wave
transmission experiments \cite{shibauchi}.  Thus, the gap structure of
$\kappa$-(ET)$_2$X salts is far from settled and the situation
strongly confronts us with the need for a powerful directional probe
of the superconducting gap.
	
During the past few years, the understanding of the heat transport in
the mixed state of superconductors with anisotropic gap has largely
progressed \cite{vekhter2}.  In particular, it was demonstrated both
experimentally and theoretically that the thermal conductivity is a
powerful tool for probing the anisotropic gap structure
\cite{yu,aubin,maki,won,vekhter1,izawa1,izawa2}.  Thermal conductivity
has some advantages, compared to other experiments.  First, it is an
unique transport quantity which does not vanish in the superconducting
state, responding to the unpaired quasiparticles (QPs).  Second, it is
a probe of the {\it bulk} free from the surface effect.  Third and
most importantly, it is indeed a {\it directional} probe, sensitive to
the relative orientation among the thermal flow, the magnetic field,
and nodal directions of the order parameter, as we will discuss in
detail later.  In fact, a clear fourfold modulation of the in-plane
thermal conductivity $\kappa$ with an in-plane magnetic field which
reflects the angular position of nodes of $d_{x^2-y^2}$ symmetry was
observed in YBa$_2$Cu$_3$O$_{7-\delta}$ \cite{yu,aubin} and 2D heavy
fermion superconductor CeCoIn$_5$ \cite{izawa2}, while such a
modulation was absent in Nb and the B-phase of UPt$_{3}$ with an
isotropic gap in the basal plane \cite{aubin,isotropic}.  These fact
demonstrate that the thermal conductivity tensor can be a relevant
probe of the superconducting gap structure.  In this Letter, we have
measured the thermal conductivity tensor of
$\kappa$-(ET)$_2$Cu(NCS)$_2$ in magnetic field rotating within the 2D
superconducting planes.  The superconducting gap structure was
successfully determined by the angular variation of $\kappa$.  On the
basis of these findings, we discuss the nature of the
superconductivity of $\kappa$-(ET)$_2$Cu(NCS)$_2$.
 
Single crystals $\kappa$-(ET)$_2$Cu(NCS)$_2$ were grown by
conventional electrochemical method and their approximate sizes are
2x1x0.1mm$^3$.  The thermal conductivity was measured by the
steady-stated method with one heater and two RuO$_{2}$ thermometers. 
The heat current {\boldmath $q$} was applied along the {\boldmath 
$b$}-direction.  The upper inset of Fig.~1 shows the FS.  In
the present measurements, it is very important to rotate {\boldmath
$H$} within the 2D {\it bc}-planes with very high accuracy because a
slight field-misalignment produces 2D pancake vortices which might act
as a strong scattering center of the thermal current.  Special care
was therefore taken to keep the perpendicular field due to the
misalignment less than $H_{c1}$ perpendicular to the layers, so that
{\it the condition for the absence of pancake vortices (lock-in state)
was always fulfilled} while rotating {\boldmath $H$} \cite{mansky}. 
For this purpose, we used a system with two superconducting magnets
generating {\boldmath $H$} in two mutually 
\begin{figure}
\centerline{\epsfxsize 7.5cm \epsfbox{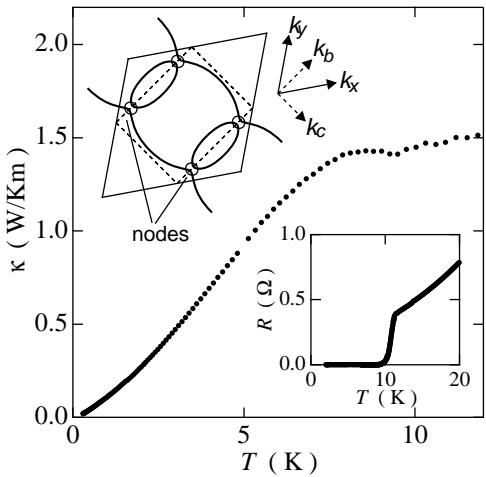}}
\caption{$T$-dependence of the thermal conductivity in zero field. 
The heat current {\boldmath $q$} was applied along the {\boldmath 
$b$}-direction.  Lower inset: The resistive transition at $T_c$. Upper
inset: The Fermi surface of $\kappa$-(ET)$_2$Cu(NCS)$_2$ (solid
lines).  The Fermi surface consists of quasi-1D and 2D hole pocket. 
The dashed lines show the first Brillouin zone with $k_{b}$ and
$k_{c}$ axes.  The thin solid lines show the extended Brillouin zone
with $k_{x}$ and $k_{y}$ axes in the similar coordinate style of the
high-$T_{c}$ cuprates.  The node directions determined in our
experiment are also shown. }
\end{figure}
\noindent orthogonal directions and a
$^{3}$He cryostat equipped on a mechanical rotating stage with a
minimum step of 1/500 degree at the top of the Dewar. 
Computer-controlling two magnets and the rotating stage, we were able
to rotate {\boldmath $H$} continuously within the 2D planes with a
misalignment of less than 0.01 degree from the plane, which we
confirmed by the simultaneous measurement of the resistivity.

We first discuss the $T$- and $H$- dependence of $\kappa$.  The
observed $T$- and $H$- dependence were very similar to the results of
Ref.\cite{belin}.  Figure~1 depicts the $T$-dependence of $\kappa$. 
Upon entering the superconducting state, $\kappa$ exhibits a kink and
rises to the maximum value just below $T_c$.  As discussed in detail
in Ref.\cite{belin}, the enhancement of $\kappa$ below $T_{c}$
reflects the increase of the phonon mean free path by the electron
condensation, which is so because the phonon thermal conductivity
$\kappa^{ph}$ well dominates over the electronic thermal conductivity
$\kappa^{el}$ near $T_c$.  Figures~2 (a) and (b) depict the
$H$-dependence of $\kappa$ in perpendicular ({\boldmath $H$}$\perp
bc-$plane) and parallel ({\boldmath $H$}$\parallel bc-$plane) field,
respectively.  In perpendicular field, $\kappa(H)$ shows a monotonic
decrease up to $H_{c2}$ above 1.6~K, which can be attributed to the
suppression of the phonon mean free path by the introduction of the
vortices.  Below 1.6~K, $\kappa(H)$  exhibits a dip below $H_{c2}$. 
The minimum of $\kappa(H)$ appears 
\begin{figure}
    \centerline{\epsfxsize 7.5cm \epsfbox{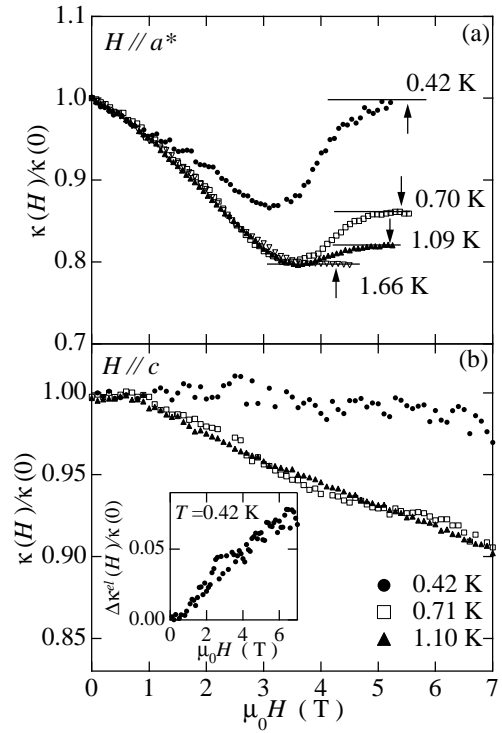}}
\caption{$H$-dependence of the in-plane thermal conductivity (a) in
perpendicular and (b) in parallel field ({\boldmath $H$} $\parallel
c$) at low temperatures.  Deviation from the horizontal line shown by
arrows marks $H_{c2}$ .  Inset: $H$-dependence of $\Delta 
\kappa^{el}/\kappa(0)$ in parallel field.  For
details, see the text.}
\end{figure}
\noindent as a result of a competition
between $\kappa^{ph}$ which always decreases with $H$ and
$\kappa^{el}$ which increases steeply near $H_{c2}$.  Then the
magnitude of the increase of $\kappa(H)$ below $H_{c2}$ provides a
lower limit of the electronic contribution.  As seen in Fig.~2(a), the
electronic contribution grows rapidly below 0.7~K;
$\kappa^{el}_n/\kappa_n$ is roughly estimated to be $\agt$5\% at 0.7~K
and $\agt$15\% at 0.42~K, where $\kappa^{el}_n$ and $\kappa_n$ are the
electronic and total thermal conductivity in the normal state above
$H_{c2}$, respectively.  This dramatic increase of
$\kappa^{el}_n/\kappa_n$ is caused by $\kappa^{ph}$ which decreases
much faster than $\kappa^{el}$ with decreasing $T$.  In parallel field
with much higher $H_{c2}$ ($\agt$30~T), $\kappa(H)$ decreases
monotonically at all temperatures.  While $\kappa(H)/\kappa(0)$ shows
a similar $H$-dependence at 0.71~K and 1.1~K, it deviates from this
pattern at 0.42~K. Since the electronic contribution grows rapidly
below 0.7~K, this deviation can be attributed to $\kappa^{el}$.  In
the inset of Fig.~2(b), we show $\Delta \kappa^{el}(H)/\kappa(0)
\left( \equiv\frac{\kappa^{el}(H)-\kappa^{el}(0)}{\kappa(0)} \right)$
at 0.4~K, assuming that $\kappa^{ph}/\kappa(0)$ has the same
$H$-dependence.

We now move on to the angular variation of $\kappa$ as {\boldmath $H$}
is rotated within the 2D planes.  Figures 3 (a)-(c) display
$\kappa$({\boldmath $H$}, $\theta)$ as a function of
$\theta=$({\boldmath $q$}, {\boldmath $H$}) at low temperatures, which
are measured in rotating $\theta$ after field cooling at
$\theta=0^{\circ}$ ({\boldmath $H$}$\parallel$ {\boldmath $b$}).  The
consecutive measurement with an inverted rotating direction did not
produce any hysteresis in $\kappa$({\boldmath $H$}, $\theta)$, which
demonstrate that the field trapping related to the pinning of the
Josephson vortices is negligibly small.  At 0.72~K,
$\kappa(${\boldmath $H$}, $\theta)$ shows a minimum at

\begin{figure}
    \centerline{\epsfxsize 7.5cm \epsfbox{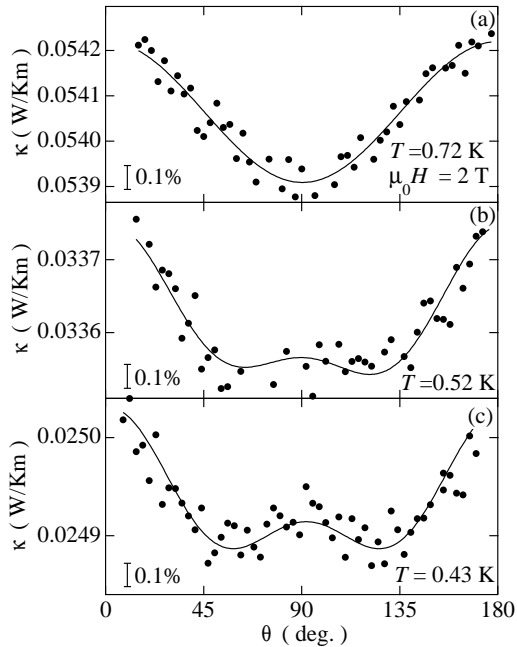}}
\caption{(a)-(c)Angular variation of $\kappa(${\boldmath
$H$},$\theta)$ in $|\mu_{0}${\boldmath $H$}$|$=2~T for different
temperatures.  $\theta=$({\boldmath $q$},{\boldmath $H$}).  The solid
lines represent the results of the fitting by the function
$\kappa(H,\theta) = C_0 +C_{2\theta}\cos2\theta +
C_{4\theta}\cos4\theta$, where $C_0$, $C_{2\theta}$ and $C_{4\theta}$
are constants.  }
\end{figure}
\noindent $\theta=90^{\circ}$.  Similar $\theta$-dependence was observed at
higher temperatures.  On the other hand, the angular variation changes
dramatically at lower temperatures, exhibiting a double minimum as
shown in Figs.~3(b) and (c).  In all data, as shown by the solid lines
in Figs.~3 (a)-(c), $\kappa$({\boldmath $H$}, $\theta)$ can be
decomposed into three terms with different symmetries; $\kappa(\theta)
= \kappa_{0} + \kappa_{2\theta} + \kappa_{4\theta}$ where $\kappa_{0}$
is a $\theta$-independent term, and $\kappa_{2\theta} =
C_{2\theta}\cos 2\theta$ and $\kappa_{4\theta} = C_{4\theta}\cos
4\theta$ are terms with two and fourfold symmetry with respect to the
in-plane rotation, respectively.  The term $\kappa_{2\theta}$, which
has a minimum at {\boldmath $H$}$\perp${\boldmath $q$}, appears as a
result of the difference of the scattering rate for QPs and phonons
traveling parallel to the vortex and for those moving in the
perpendicular direction.  Since a large twofold symmetry is observed
even above 0.7~K where $\kappa^{ph}$ dominates, $\kappa_{2\theta}$ is
mainly phononic in origin.  In what follows, we will address the
fourfold symmetry which is directly related to the electronic
structure.
	
Figures 4 (a)-(c) display $\kappa_{4\theta}$ normalized by $\kappa_n$
after the subtraction of the $\kappa_0$- and $\kappa_{2\theta}$-terms
from the total $\kappa$.  At $T$=0.72~K, the fourfold component is
extremely small; $|C_{4\theta}|/\kappa_n<0.1$\%.  On the other hand, a
clear fourfold component with $|C_{4\theta}|/\kappa_n\sim 0.2$\% is
resolved at 0.52 and 0.43~K. Since the contribution of $\kappa^{el}$
grows rapidly below 0.7~K and occupies a substantial portion of the
total $\kappa$ at 0.4~K, it is natural to consider that {\it the
fourfold oscillation is purely electronic in origin.} Although
$|C_{4\theta}|$ at 0.42~K is as small as 0.2\% in $\kappa_n$, it
occupies approximately 1.5-2\% in $\kappa^{el}_n$ and occupies a few
\% in $\kappa^{el}(0)$ assuming $\kappa^{el}_n/\kappa_n\sim$
\begin{figure}
    \centerline{\epsfxsize 7.1cm \epsfbox{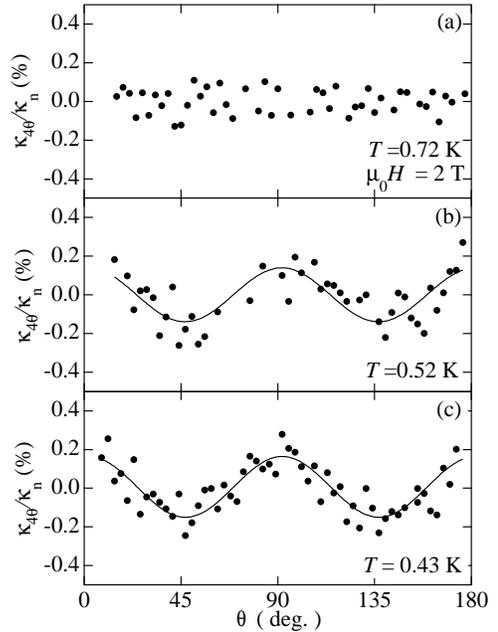}}
\caption{(a)-(c) The fourfold symmetry $\kappa_{4\theta}$ obtained 
from Figs.~3 (a)-(c).  The solid lines represent $C_{4\theta}\cos 
4\theta$. For details,  see the text. }
\end{figure}
\noindent 0.15. We
note that this value is one order of magnitude larger than that in 
Sr$_2$RuO$_4$ with an isotropic gap in the plane \cite{izawa1}.  We
now  address the origin for the fourfold symmetry.  The most important
issue here is "Is the observed fourfold symmetry in $\kappa^{el}$ a
consequence of the line nodes perpendicular to the layer?".  We will
show that the band structure inherent to the crystal is very unlikely
to be an origin of the fourfold symmetry.  First of all, it can be
shown by the group theoretical consideration that the anisotropic term
in $\kappa^{el}$ due to a fourfold distortion of the FS is of second
order relative to the leading terms, since the thermal conductivity
$\kappa_{xx}$ is a second rank tensor \cite{maki2}.  In addition, the
crystal structure of $\kappa$-(ET)$_2$Cu(NCS)$_2$ is monoclinic and FS
is nearly elliptic; the fourfold distortion of the FS should be very
small if it exists.  Second, the in-plane magnetoconductivity $\Delta
\sigma(${\boldmath $H$}$)=\sigma(${\boldmath $H$}$)-\sigma(H=0)$ above
$T_c$ is undetectably small even at 5~T due to the very strong two
dimensionality.  In fact, the upper limit of $\Delta \sigma/\sigma(0)$
roughly estimated from the warp of the FS perpendicular to the plane
is less than 10$^{-5}$ at 2~T. Thus, as far as the Wiedemann-Franz law
holds, the fourfold oscillation of $\kappa_{el}$ arising from the
magnetoconductance should be undetectably small.  These considerations
lead us to conclude that {\it the observed fourfold symmetry
originates from the superconducting gap nodes} \cite{swave}.
		
In the thermal transport in the superconductors with nodes, the
dominant effect in a magnetic field is the Doppler shift of the
delocalized QP energy spectrum, which occurs due to the presence of a
superfluid flow around each vortex, and generates a nonzero QP density
of states (DOS) at the Fermi level (Volovik effect) \cite{volovik}. 
While the Doppler shift increases $\kappa^{el}$ with $H$ through the
enhancement of the DOS, it can also lead to a decrease of
$\kappa^{el}$ through the suppression of impurity scattering time and
Andreev scattering time off the vortices.  At high temperatures, the
latter effect is predominant, but eventually gives way to the former
at low temperatures, as demonstrated in high-$T_c$ cuprates
\cite{chiao}.  Since $\kappa^{el}$ increases with $H$ as shown in the
inset of Fig.~2(b), the enhancement of the DOS is the main origin for
the $H$-dependence of $\kappa^{el}$ at 0.42~K \cite{microwave}.  In
this case, rotating {\boldmath $H$} within the 2D-plane gives rise to
the fourfold oscillation in $\kappa^{el}$ associated with the DOS
oscillation \cite{won,vekhter1}.  This effect arises because the DOS
depends sensitively on the angle between {\boldmath $H$} and the
direction of the nodes of the order parameter, because the QPs
contribute to the DOS when their Doppler-shifted energies exceed the
local energy gap.  The DOS oscillation with fourfold symmetry gives
rise to $\kappa^{el}$ which attains its maximum value when {\boldmath
$H$} is directed to the antinodal directions and turns minimum when
{\boldmath $H$} is directed along the nodal directions (see Fig.~2 in
Ref.\cite{vekhter1}).  According to Ref.\cite{won}, the amplitude of
the fourfold symmetry in the $d$-wave superconductors arising from the
DOS oscillation is roughly estimated as $|C_{4\theta}|/\kappa^{el}(0)
= 0.082\frac{\upsilon_F\upsilon^{\prime}_F
eH}{3\pi\Gamma\Delta}\ln(\sqrt{32\Delta/\pi\hbar\Gamma})$. 
Here $\Delta$ is the superconducting gap, $\Gamma$ is the QP
relaxation rate, $\upsilon_F$ and $\upsilon^{\prime}_F$ are the
in-plane and out-of-plane Fermi velocity, respectively.  Using
$\Gamma\sim2\times10^{11}$s$^{-1}$, $2\Delta/k_BT_c$=3.54,
$\upsilon_F\sim5\times10^4$m/s, and $\upsilon^{\prime}_F
\sim2.5\times10^3$m/s, gives $|C_{4\theta}|/\kappa^{el}(0)\sim$3\%. 
Thus the DOS oscillation by Volovik effect yields
$|C_{4\theta}|/\kappa^{el}(0)$ which is in the same order to the data.
	 
The fourfold symmetry enables us to specify the node directions, which
is crucial for understanding the pairing interaction. 
$\kappa_{4\theta}$ exhibits a maximum when {\boldmath $H$} is applied
parallel to the {\boldmath $b$} and {\boldmath $c$} axes of the
crystal, showing {\it the gap nodes along the directions rotated
45$^{\circ}$ relative to the {\boldmath $b$} and {\boldmath
$c$}-axes}; the nodes are situated near the band gap between the 1D
and 2D bands (see the upper inset of Fig.~1).  This result is
consistent with the STM experiments \cite{nomura}.  We emphasize here
that {\it the determined nodal structure is inconsistent with the
recent theories based on the AF fluctuation}.  If one assumes an AF
fluctuation scenario, it is natural to expect the nodes to be along
the {\boldmath $b$} and {\boldmath $c$} directions.  This is because
the AF ordering vectors become parallel to the {\boldmath $b$}-axis,
which would provide a partial nesting.  If we take the same
conventions for the magnetic Brillouin zone as the high-$T_c$ cuprates
with $d_{x^2-y^2}$ symmetry (see Fig.~1 (c) in Ref.\cite{AFSF1}), the
superconducting gap symmetry of $\kappa$-(ET)$_2$Cu(NCS)$_2$ is
$d_{xy}$.  It is interesting to note that superconductivity with
$d_{xy}$ symmetry has been theoretically suggested based on the charge
fluctuation scenario \cite{mckenzie2}.  Our present results may bear
implications on this issue.

We finally comment on the recent heat capacity measurements which
report a fully gaped superconductivity \cite{elsinger}.  In our view,
their temperature range ($T>T_c/5$) is not low enough to conclude the
exponential behavior of the heat capacity; the measurements at
temperatures less than $T_c$/10 would be required.
	
In summary, the thermal conductivity tensor of
$\kappa$-(BEDT-TTF)$_2$Cu(NCS)$_2$ was studied in a magnetic field
rotating within the 2D superconducting planes.  The observed fourfold
oscillation provides a strong evidence of $d$-wave symmetry.  From its
sign, the node directions are successfully specified.  These results
place strong constraints on models that attempt to explain the
mechanism of the superconductivity of
$\kappa$-(BEDT-TTF)$_2$Cu(NCS)$_2$.
	 
We thank M.~Imada, K.~Kanoda, K.~Kuroki, M.~Ogata, A.~Tanaka and
K.~Ueda for stimulating discussions.  We are also grateful to K.~Maki
for several comments and for showing unpublished results.



\begin{thebibliography}{99}
\bibitem{kanoda}K.~Kanoda, Physica C {\bf 282-287}, 299 (1997) and 
references therein.
\bibitem{mckenzie1}For review, see R. H.~McKenzie, Science {\bf 278}, 
820 (1997);  Comments Cond. Matt. Phys. {\bf 18}, 309 (1998).
\bibitem{AFSF1}J. Schmalian, Phys. Rev. Lett. {\bf 81}, 4232 (1998). 
\bibitem{AFSF2}H.~Kino, and H.~Kontani, J. Phys. Soc. Jpn. {\bf 67},  
3691 (1998); H.~Kondo, and T.~Moriya, {\it ibid.} {\bf 67}, 3695 
(1998); K.~Kuroki and H.~Aoki, Phys. Rev. B {\bf 60},  3060 (1999). 
\bibitem{jerome}H.~Mayaffre {\it et al.}, Phys Rev. Lett. {\bf 75}, 
4122 (1995); S. D. De Soto {\it et al.} Phys. Rev. B {\bf 52}, 10364 
(1995).
\bibitem{belin}S.~Belin {\it et al.}, Phys. Rev. Lett. {\bf 81}, 4728 
(1998).
\bibitem{elsinger}H.~Elsinger {\it et al.}, Phys Rev. Lett. {\bf 84}, 
6098 (2000); J.~M\"{u}ller {\it et al.}, cond-mat/0109030.
\bibitem{nomura}T.~Arai {\it et al.}, Phys. Rev. B {\bf 63}, 104518 
(2001).
\bibitem{schrama}J.M.~Schrama {\it et al.}, Phys. Rev. Lett. {\bf 
83}, 
3041 (1999).
\bibitem{shibauchi}S.~Hill  {\it et al.}, Phys. Rev. Lett. {\bf 86}, 
3451 (2001); T.~Shibauchi {\it et al.}, {\it ibid} {\bf 86}, 3452 
(2001).
\bibitem{vekhter2}C.~K\"ubert and P.J.~Hirschfeld, Phys.  Rev.  Lett. 
{\bf 80}, 4963 (1998); M.~Franz, {\it ibid.} {\bf 82}, 1760 (1999); 
I.~Vekhter and A.~Houghton, {\it ibid.} {\bf 83}, 4626 (1999); 
Yu.S.~Barash and A. A.~Svidzinsky, Phys.  Rev.  B {\bf 58}, 6476 
(1998).
\bibitem{yu}F.~Yu {\it et al.}, Phys. Rev. Lett. {\bf 74}, 5136 
(1995). 
\bibitem{aubin} H.~Aubin {\it et al.}, Phys. Rev. Lett. {\bf 78}, 
2624 
(1997). 
\bibitem{maki}K.~Maki {\it et al.}, Physica C {\bf 341-348}, 1647 
(2000).
\bibitem{won}H.~Won and K.~Maki, cond-mat/0004105. 
\bibitem{vekhter1}I.~Vekhter {\it et al.}, Phys.  Rev.  B {\bf 59}, 
R9023 (1999).  
\bibitem{izawa1}K.~Izawa {\it et al.}, Phys. Rev. Lett. {\bf 86}, 2653 
(2001).
\bibitem{izawa2}K.~Izawa {\it et al.}, Phys. Rev. Lett. {\bf 87}, 
057002 (2001). 
\bibitem{isotropic} J.~Lowell and J.B.~Sousa, J. Low Temp.  Phys. 
{\bf 3}, 65 (1970); H.~Suderow {\it et al.}, Phys.  Lett.  A {\bf 
234},
64 (1997).
\bibitem{mansky}P.A.~Mansky {\it et al.}, Phys Rev. B {\bf 50}, 15929 
(1994).	  
\bibitem{maki2}K.~Maki, private commun.
\bibitem{swave} We comment on a very anisotropic $s$-wave state.  The
lower limit of the gap anisotropy ratio $\Delta_{max}/\Delta_{min}$ is
roughly estimated from the observed fourfold oscillation as
$\Delta_{max}/\Delta_{min}\gg
\Delta_{max}/\sqrt{\hbar\upsilon_{F}\upsilon'_{F}eH}\simeq 5$. 
However an $s$-wave gap with such a large anisotropy is very unlikely
to be realized in $\kappa$-(ET)$_2$Cu(NCS)$_2$ with its rather simple
FS.
\bibitem{volovik}G.E.~Volovik, JETP Lett.  {\bf 58}, 469 (1993).
\bibitem{chiao}H.~Aubin {\it et al.}, Phys. Rev. Lett. {\bf 82}, 624 
(1999); M.~Chiao {\it et al.}, {\it ibid.} {\bf 82}, 2943 (1999).
\bibitem{microwave}  The microwave surface impedance measurements 
show that the suppression of the QP scattering rate below $T_c$ is 
absent.  This fact implies that the Andreev scattering is much less 
important for $\kappa(H)$ compared to  high-$T_c$ cuprates  and gives 
support to our conclusion.	
\bibitem{mckenzie2} D.J.~Scalapino {\it et al.}, Phys. Rev. B,{\bf 
35}, 6694 (1987); J.~Merino and R.H.~Mckenzie, cond-mat/0106425.
	  

\end{thebibliography}
\end{document}